# Expert Knowledge-guided Geometric Representation Learning for Magnetic Resonance Imaging-based Glioma Grading

Yeqi Wang, Longfei Li, *Student, IEEE*, Cheng Li, Yan Xi, Hairong Zheng, *Senior Member, IEEE,* Yusong Lin, *Member, IEEE,* Shanshan Wang\*, *Senior Member, IEEE*

*Abstract*—Radiomics and deep learning have shown high popularity in automatic glioma grading. Radiomics can extract hand-crafted features that quantitatively describe the expert knowledge of glioma grades, and deep learning is powerful in extracting a large number of high-throughput features that facilitate the final classification. However, the performance of existing methods can still be improved as their complementary strengths haven't been sufficiently investigated and integrated. Furthermore, lesion maps are usually needed for the final prediction at the testing phase, which is very troublesome. In this paper, we propose an expert knowledge-guided geometric representation learning (ENROL) framework . Geometric manifolds of hand-crafted features and learned features are constructed to mine the implicit relationship between deep learning and radiomics, and therefore to dig mutual consent and essential representation for the glioma grades. With a specially designed manifold discrepancy measurement, the grading model can exploit the input image data and expert knowledge more effectively in the training phase and get rid of the requirement of lesion segmentation maps at the testing phase. The proposed framework is flexible regarding deep learning architectures to be utilized. Three different architectures have been evaluated and five models have been compared, which show that our framework can always generate promising results.

## I. INTRODUCTION

GLIOMA are the most common type of primary neuroepithelial malignant tumors with high incidence and mortality rates [1]. Gliomas account for nearly 30% of all primary brain tumors and 80% of malignant tumors [2]. According to their histological appearance, gliomas can be classified as astrocytic, oligodendroglial, and ependymal tumors. The World Health Organization (WHO) classifies gliomas into four grades (grades I-IV) according to the degree of malignancy [3]. Generally, gliomas can be referred to as low-grade gliomas (LGGs, grades I and II) and high-grade gliomas (HGGs, grades III and IV). In clinical practices, the treatment and prognosis of LGG and HGG patients vary substantially. LGG patients are mainly treated by surgery with a good prognosis, and about 70% of LGG patients can survive for more than 10 years [4]. HGG patients are usually treated by surgery combined with chemotherapy and have a poor prognosis. Only 50% of HGG patients can survive for half a year [5]. Therefore, accurate glioma grading is critical for the development of effective and personalized treatment plans.

Magnetic resonance imaging (MRI) plays an important role in the diagnosis of gliomas because of its superior imaging contrast and non-invasiveness [6]. MRI can help localize brain lesions preoperatively and evaluate the curative effects postoperatively. Nevertheless, inspecting 3D MR images manually requires rich domain knowledges and experiences. Due to the complexity of the shape and texture of gliomas, manual MRI-based diagnosis is time-consuming and error-prone. To assist physicians in achieving fast and accurate diagnoses, automatic MRI-based glioma grading models are urgently needed.

Radiomics [7] has been a widely utilized computer-aided diagnosis (CAD) method for glioma grading recently. Radiomics-based methods include four major steps: delineate

This research was partly supported by Scientific and Technical Innovation 2030-"New Generation Artificial Intelligence" Project (2020AAA0104100, 2020AAA0104105), the National Natural Science Foundation of China (61871371，81830056), Key Laboratory for Magnetic Resonance and Multimodality Imaging of Guangdong Province (2020B1212060051), the Basic Research Program of Shenzhen (JCYJ20180507182400762), Youth Innovation Promotion Association Program of Chinese Academy of Sciences (2019351).

Yeqi Wang, and Longfei Li are with the School of Computer and Artificial Intelligence, Zhengzhou University, Zhengzhou 450001, China, also with the Paul C. Lauterbur Research Center for Biomedical Imaging, Shenzhen Institutes of Advanced Technology, Chinese Academy of Sciences, Shenzhen 518055, China (e-mail: yeqiwangst@163.com ; lfli@ha.edu.cn ).

Cheng Li, Yan Xi, and Hairong Zheng are with the Paul C. Lauterbur Research Center for Biomedical Imaging, Shenzhen Institutes of Advanced Technology, Chinese Academy of Sciences, Shenzhen 518055, China (e-mail: cheng.li6@siat.ac.cn ; yanxi@first-imaging.com ; hr.zheng@siat.ac.cn ).

Yusong Lin are with the School of Software, Zhengzhou University, Zhengzhou 450002, China, also with the Collaborative Innovation Center for Internet Healthcare and the Hanwei IoT Institute, Zhengzhou University, Zhengzhou 450002, China (e-mail: yslin@ha.edu.cn ).

Shanshan Wang is with the Paul C. Lauterbur Research Center for Biomedical Imaging, Shenzhen Institutes of Advanced Technology, Chinese Academy of Sciences, Beijing 100864, China, also with the Peng Cheng Laboratory, Shenzhen 518066, China, and also with the Guangdong Provincial Key Laboratory of Artificial Intelligence in Medical Image Analysis and Application, Guangdong Provincial People's Hospital, Guangdong Academy of Medical Sciences, Guangzhou 510080, China (e-mail: sophiasswang@hotmail.com).

Yeqi Wang and Longfei Li contribute equally to the article.

the lesions in brain MR images, extract quantitative hand-crafted features from the segmented lesion regions, select the key features, and train a classifier with the selected features. The hand-crafted features mainly describe the statistics, shape, and texture information of the lesions, which are sometimes difficult for physicians to observe with naked eyes directly. These features have shown great applicability in glioma grading tasks. For example, Zhao et al. [8] conducted a retrospective study with 400 glioma patients. To reduce the efforts on lesion delineation, they firstly compressed 3D MR images into 2D images using the max pooling operation. The selected key features included first order statistics and textures, and then, a support vector machine (SVM) classifier was trained. Their method achieved an AUC of 0.902 and a sensitivity of 0.8604 for glioma grading. Cho et al. [9] employed the open-source dataset BraTs2018 (285 patients) in their study. They found that shape and texture features were more important for glioma grading. Three different classifiers were trained, including random forest, SVM, and logistic regression. They achieved an average AUC of 0.9030 and an average accuracy of 0.8854. Vamvakas et al. [10] investigated the significance of MR imaging biomarkers (built by extracted quantitative features) on glioma grading utilizing the data from 40 glioma patients. The key quantitative features were 3D texture descriptors calculated from the histogram gray level co-occurrence matrix (GLCM), and gray level run length matrix (GLRM). The final imaging biomarkers were developed by training a linear sequential minimal optimization (SMO) classifier with the key features. They obtained an AUC of 0.955, an accuracy of 0.955, and a sensitivity of 0.95, and validated that the imaging biomarkers were important for the glioma grading task. Tian et al. [11] performed a retrospective study with data from 153 glioma patients. They extracted texture features from the lesion regions and trained the SVM-based glioma grading model with a classification AUC of 0.987 and an accuracy of 0.968. Similarly, Su et al. [12] conducted a retrospective study with data from 220 glioma patients. They used a semi-automatic method to assist physicians in delineating the lesion regions, and extracted key features to train a multivariate model. The final grading model achieved good results with an AUC of 0.911. These studies validated that radiomics-based methods can achieve encouraging glioma grading results on the respective datasets and the manually extracted quantitative features play important roles in the development of the grading models. Nevertheless, the heavy reliance on hand-crafted features reduces the universality and generalizability of the developed models.

Recently, deep learning, especially convolutional neural networks (CNNs), has presented unprecedented performance in many computer vision tasks, such as classification [13], segmentation [14], and detection [15]. Similarly, CNNs have also been widely employed in medical imaging. Different from radiomics, CNNs can automatically extract high-dimensional features from inputs, and they can accomplish tasks in an end-to-end manner without the prerequisite of the expert knowledge about the glioma grades. For glioma grading, there are a lot of studies adopting CNN-based methods. Matsui et al. [16] developed a multi-input 2D CNN model to make better use of the training data. The inputs consisted of the whole brain MR images, the tumor regions, and extract clinical information of the patients. By utilizing both imaging data and the clinical information, they achieved relatively good results on the prediction of LGG molecular subtypes with an accuracy of 0.687. Sultan et al. [17] developed a 2D CNN model to address the task of multi-classification of brain tumors. T1 weighted with contrast enhanced (T1ce) MR images were used, and the authors evaluated the effectiveness of their method on two open-source datasets. They achieved the best overall accuracy of 0.9613 and 0.9870 on two tasks of multi-classification of brain tumors. One is to classify tumors into meningioma, glioma, and pituitary tumor, and the other is to differentiate grade II, grade III, and grade IV gliomas. Yang et al. [18] investigated the effectiveness of transfer learning on glioma grading. They utilized 2D MR image slices from 113 patients to finetune ImageNet-pretrained AlexNet and GoogLeNet. Compared to the models trained from scratch, the transferred models showed better grading performance with an AUC of 0.968 and an accuracy of 0.945. Similarly, Deepak et al. [19] applied transfer learning to the brain tumor multi-classification task. They finetuned the ImageNet-pretrained GoogLeNet by freezing the feature extraction layers. Then, they built a k-nearest neighbors (KNN) classifier utilizing the extracted features and achieved a classification accuracy of 0.98. Ge et al. [20] developed a multi-scale 3D CNN model for glioma grading. They fused the multi-scale features extracted by different network layers in the model through the utilization of skip connections. The model achieved a good grading performance with an accuracy 0.8947. Following this work, Ge et al. [21] further built a multi-stream 2D CNN model to extract and fuse the features from multi-parametric MR images for glioma grading. This method obtained a good glioma grading performance with an accuracy of 0.9087.

CNNs contain numerous parameters and complex structures to build a large hypothesis space, which contribute to the powerful fitting capability of CNN-based models. These studies have made encouraging progresses for glioma grading. However, there are still some essential challenges that remain to be solved. It is very challenging to search the optimal hypothesis of CNNs with insufficient training data in medical imaging tasks. To alleviate this problem, existing studies typically adopt lesion segmentation maps and took lesion regions as inputs, which reduces the complexity of the hypothesis space and eases the fitting process of CNN-based models. However, this makes it necessary to delineate the lesion regions first in the clinical use of the developed grading models. Manually delineating the brain lesions regions in 3D MR images is time-consuming and resource-intensive. Moreover, different physicians may give slightly different segmentation results, which can affect the grading performance during model testing. To this end, automatic brain lesion segmentation algorithms have been developed to serve as a first step in the glioma grading process [22], [23]. The segmentation performance is crucial for the subsequent glioma grading, but automatic and accurate brain lesion segmentation on its own is a difficult task. Therefore, there is an urgent need for the development of high-performance and physician-friendly glioma grading models, which can achieve the fusion with the expert knowledge that describe the information of lesion regions in the training phase and the elimination reliance on lesion segmentation maps in the testing phase.

Some studies [24], [25] have introduced radiomics into deep learning, and merged the radiomic features with the deep learning features. Although this method has limited improvement, it shows the effectiveness of radiomics for improving the performance of deep learning. In this paper, we propose an expert knowledge-guided geometric representation learning (ENROL) framework for glioma grading without the lesion segmentation requirements at the testing phase. With the ENROL framework, glioma grading models can extract the mutual consent representations between the expert knowledge based radiomics features and data-driven deep learning features that describe the variations of gliomas of different grades. Two major steps are involved. The first step is to construct geometric manifolds in a joint latent space. Quantitative hand-crafted features are extracted from lesion regions in brain MR images. These features describe specific attributes of lesion regions. Then, high-dimensional learned features are extracted by a specific feature extractor from the whole MR images. Due to the inconsistence of the two feature spaces, encoders are developed to map these features into a joint latent space, and two geometric manifolds are constructed respectively. The second step is to train a glioma grading model with our specially designed manifold regularization item. A manifold discrepancy distance that measures the distance between the constructed manifolds is designed to regularize the grading model training. Expert knowledge is introduced to the parameter optimization of the model through minimizing the manifold discrepancy distance. Here, the expert knowledge refers to the attributes of the lesion regions embedded in the hand-crafted features. In this way, the grading model can exploit MR images more effectively without the help of lesion segmentation maps in testing. A high-performance and physician-friendly grading model can then be obtained for testing without the need of segmentation maps.

Specifically, the contributions of the proposed ENROL framework are three-fold:

1) Geometric manifolds of hand-crafted features and learned features are constructed to mine the implicit relationship between deep learning and radiomics, and therefore to dig mutual consent and essential representation for the glioma grades.

2) A high-performance and physician-friendly model is developed for glioma grading. With a deliberately designed manifold discrepancy measurement, the grading model can exploit the input MR image data and the expert knowledge more effectively during training and get rid of the requirement of lesion segmentation maps at the testing phase.

3) The proposed ENROL framework is highly flexible regarding the specific feature extraction networks to be utilized.

We have evaluated our framework with three different architectures (VGG, ResNet, and SEResNet) on four models in total compared, which show our framework can give promising glioma grading results.

## II. METHOD

### A. Dataset

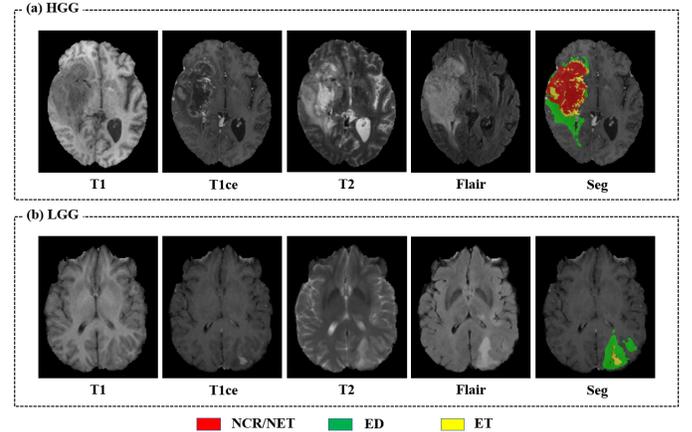

**Fig. 1.** Example brain MR images. The delineated brain lesions contain the necrotic and non-enhancing tumor core (NCR/NET), the peritumoral edema (ED), and GE-enhancing tumor (ET). The whole lesion regions, which comprise NCR/NET, ED and ET, are used to extract the quantitative hand-crafted features in the proposed ENROL framework.

The Brain Tumor Segmentation Challenge 2020 (BraTS 2020) dataset [26] is utilized in this study. The dataset contains 369 glioma samples, with 293 HGG samples and 76 LGG samples. The samples are randomly divided into a training dataset and a test dataset with a ratio of 4:1. The training dataset contains 296 samples (235 HGG samples and 61 LGG samples), and the test dataset contains 73 samples (58 HGG samples and 15 LGG samples). All of the samples include four sequences, including T1 weighted, T1ce, T2 weighted, and fluid-attenuated inversion recovery (FLAIR) (Fig. 1). Brain lesions in the MR images have been segmented manually following the same annotation protocol. The segmentation results have been approved by experienced neuro-radiologists. In addition, all the MR images have been properly pre-processed. They have been co-registered to the same anatomical template, interpolated to the same resolution ($1mm^3$) in the standard axial direction, and skull-stripped.

### B. Training Paradigm of the Glioma Grading Model

In this study, our task is to build a deep learning model for glioma grading. Let $s = (x, y) \in \mathcal{X} \times \mathcal{Y}$ be a training sample pair, where $x$ represents a brain MR image, and $y$ is the corresponding glioma grading label. We try to learn a nonlinear deep learning model (DLM) to classify x to the type of y. The hypothesis of the grading neural network $h_{dlm}$ is a function which belongs to the hypothesis class $\mathcal{H}_{dlm}$, namely $h_{dlm} \in \mathcal{H}_{dlm}$. A loss function $\ell: h_{dlm}(\mathcal{X}) \times \mathcal{Y} \rightarrow \mathbb{R}_+$ is proposed, where $\ell(h_{dlm}(x), y)$ is the loss value of $h_{dlm}$ with the sample $s = (x, y)$. Given $\mathcal{D}_{train} = \{s_i = (x_i, y_i)\}_{i=1}^N$ as a set of

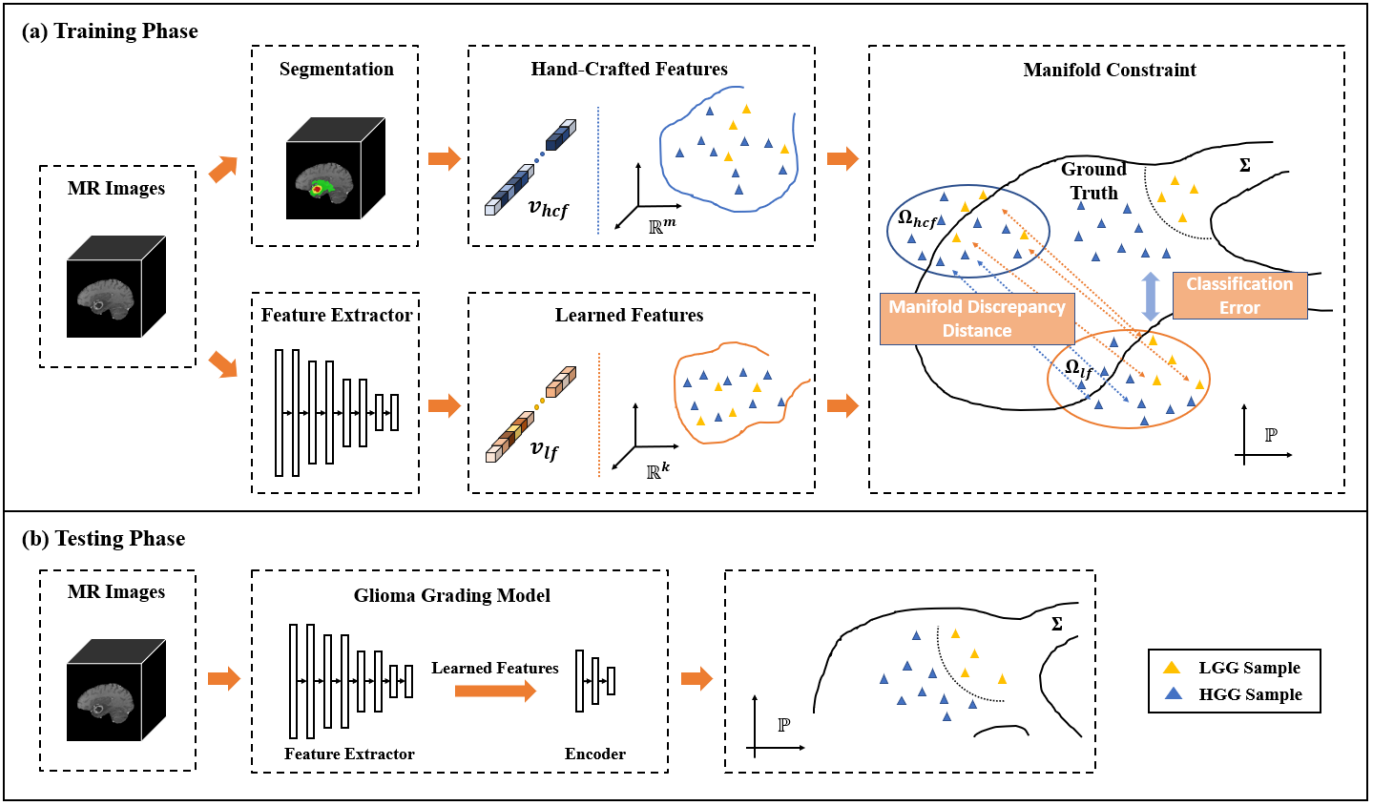

**Fig. 2.** Schematic of the proposed ENROL framework for glioma grading. At the training phase, hand-crafted features are firstly extracted from the MR images, which can describe the attributes of the lesion regions. High-dimensional learned features are extracted by a feature extractor. Then, encoders are designed to constructed the geometric manifolds $\Omega_{lcf}$ and $\Omega_{lf}$ in the joint latent space $\mathbb{P}$ with the hand-crafted features and learned features as inputs, respectively. Here, $\Omega_{lcf}$ contains the expert knowledge information of the lesion regions. To force the extractor utilizing the image data more effectively in training, a manifold-discrepancy distance is calculated and minimized. At the testing phase, the feature extractor and encoder trained with the proposed ENROL framework can be tested as a glioma grading model without requiring the inputs of brain lesion segmentation maps in the testing phase.

training samples, the empirical loss of the grading neural network on $\mathcal{D}_{train}$ is defined as:

$$\mathcal{L}_{train}(\hbar_{dlm}) = \frac{1}{N}\sum_{i=1}^{N}\ell(\hbar_{dlm}(x_i), y_i).$$

Hence, the grading neural network is trained on the training dataset $\mathcal{D}_{train}$, which aims to optimize the following objective function:

$$\hbar_{dlm}^{*}(x) = \underset{\hbar_{dlm}\in\mathcal{H}_{dlm}}{argmin}\ \mathcal{L}_{train}(\hbar_{dlm}) + \lambda\mathcal{R}(\hbar_{dlm})$$

$$= \underset{\hbar_{dlm}\in\mathcal{H}_{dlm}}{argmin}\ \frac{1}{N}\sum_{i=1}^{N}\ell(\hbar_{dlm}(x_i), y_i) + \lambda\mathcal{R}(\hbar_{dlm}),$$

where $\mathcal{R}(\hbar_{dlm})$ is a regularization item to modulate the hypothesis function $\hbar_{dlm}$ and $\lambda$ is a regularization parameter.

### C. Expert Knowledge-guided Geometric Representation Learning

Our proposed ENROL framework treats hand-crafted features as expert knowledge and utilize them to facilitate the training of the grading models. Specifically, a total of 1618 quantitative features are extracted from the lesion regions using the open-source python package PyRadiomics. The extracted features can be divided into three categories, first-order statistic features, shape and size features, and texture features. They can describe the distributions of voxel intensity values, 3D shape and size attributes, and texture attributes, respectively. All the features are normalized to the range of 0 to 1. Due to the non-Euclidean structures of the hand-crafted features, it is challenging to introduce them into the deep learning models in an effective way. To address this issue, our ENROL framework utilizes a geometric representation learning framework to handle these features, achieving the sufficient fusion of the hand-crafted features and the learned features. The overall schematic is shown in Fig. 2.

In our proposed ENROL framework, the expert knowledge is used as a prior to explore the relationship between the geometric manifolds in the feature space. high-dimensional learned features are first extracted. Due to the inconsistence between the hand-crafted feature space and the learned feature space, we use encoders to map these features into a joint latent space $\mathbb{P}$, where the mapped two sets of feature points can be utilized to construct the two geometric manifolds $\Omega_{hcf}$ of hand-crafted features (HCF) and $\Omega_{lf}$ of learned features (LF), respectively. Hypothesis $\hbar_{epk}\in\mathcal{H}_{epk}$ is built to describe the construction process of $\Omega_{hcf}$, where $\mathcal{H}_{epk}$ is a hypothesis class for the expert knowledge (EPK). In the joint latent space $\mathbb{P}$, we can exploit the implicit information of inter- and intra-glioma grading class variations more comprehensively and define the discrepancy between the two feature manifolds more accurately. Here, KL-divergence is used to measure the discrepancy between $\Omega_{hcf}$ and $\Omega_{lf}$. With the minimization of KL-divergence at the training phase, the expert knowledge embedded in $\Omega_{hcf}$ can be introduced to assist the parameter optimization of the grading model.

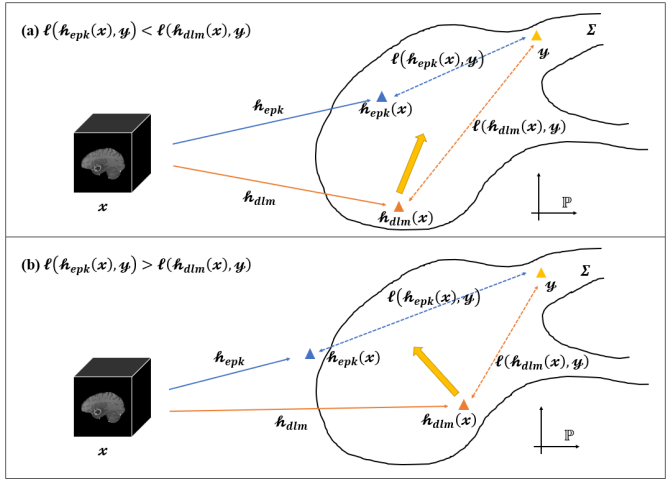

**Fig. 3.** Introducing the expert knowledge regularization item may have different effects at the training phase. Given a brain MR image s, we define the distances between the mapping points of the hypotheses $h_{epk}$ and $h_{dlm}$ and the true label by $\ell(h_{epk}(x), y)$ and $\ell(h_{dlm}(x), y)$, respectively. (a) When $\ell(h_{dlm}(x), y)$ is larger than $\ell(h_{epk}(x), y)$, the introduction of the expert knowledge regularization item can help the grading models achieve better performance. (b) When $\ell(h_{dlm}(x), y)$ is smaller than $\ell(h_{epk}(x), y)$, introducing the expert knowledge regularization item may misdirect the convergence direction of the grading models, and lead to worse performance.

The effect of the expert knowledge on the parameter optimization of the grading model with a specific sample s is illustrated in Fig. 3. Given a MR image $x$, we can get the mapping points $h_{epk}(x)$ and $h_{dlm}(x)$ in geometric manifolds $\Omega_{hcf}$ and $\Omega_{lf}$. The distances between the mapping points and the true label $y$ in $\Sigma$ are defined by $\ell(h_{epk}(x), y)$ and $\ell(h_{dlm}(x), y)$, respectively. When $\ell(h_{dlm}(x), y)$ is larger than $\ell(h_{epk}(x), y)$, the introduction of the expert knowledge embedded in $\Omega_{HCF}$ can help the grading model converge towards the optimal solution. If $\ell(h_{dlm}(x), y)$ is smaller than $\ell(h_{epk}(x), y)$, introducing the expert knowledge regulation may hinder the convergence of the grading model and lead to worse grading performance. To prevent the possible negative effect brought by the inaccurate expert knowledge introduced by radiomics, a deliberated manifold discrepancy distance is proposed. We follow two rules when designing the distance measurement:

(1) when $\ell(h_{dlm}(x), y)$ is larger than $\ell(h_{epk}(x), y)$, the manifold-discrepancy distance should increase with the difference between the two distances;

(2) when $\ell(h_{dlm}(x), y)$ is smaller than $\ell(h_{epk}(x), y)$, the manifold-discrepancy distance should be close to 0.

Therefore, the manifold-discrepancy distance is defined as:

$$Dist\langle h_{epk}(x), h_{dlm}(x)\rangle =$$
$$\sigma[\ell(h_{epk}(x), y) - \ell(h_{dlm}(x), y)]$$
$$\times KL\left(h_{epk}(x) \parallel h_{dlm}(x)\right),$$

where $KL(p \parallel q)$ refers to the KL-divergence and $\sigma(\cdot)$ is the ReLU activation. Adding the manifold-discrepancy distance, the overall objective function of the grading model is:

$$h_{dlm}^*(x) = \underset{h_{dlm}\in\mathcal{H}_{dlm}}{argmin} \frac{1}{N}\sum_{i=1}^{N}\ell(h_{dlm}(x_i), y_i)$$
$$+ \lambda Dist\langle h_{epk}(x_i), h_{dlm}(x_i)\rangle.$$

When optimizing the grading model utilizing $h_{dlm}^*(x)$, a high-performance and physician-friendly grading model is obtained, which can extract more powerful and effective representations of gliomas from MR images and achieve improved grading results without segmentation requirements.

### D. Implementation Details

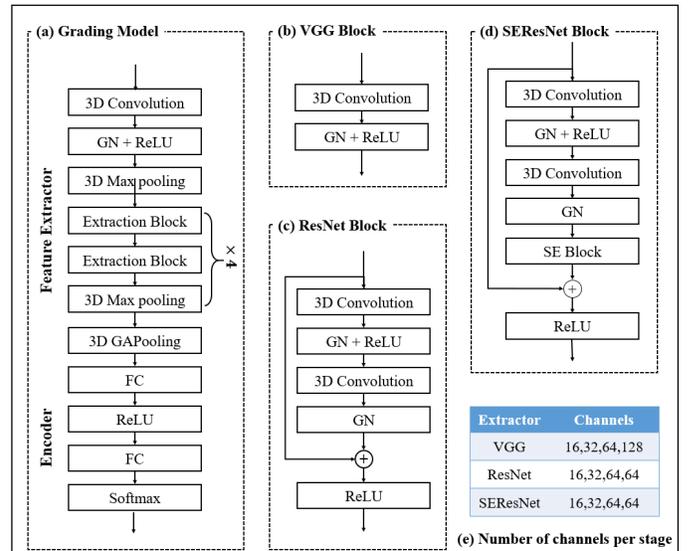

**Fig. 4.** Architectures of the constructed grading models. (a) The overall architecture of the grading models. All three models adopt a similar overall architecture, including a feature extractor part and an encoder part. The main difference lies in the design of the extraction block in the feature extractor part. (b) The VGG feature extraction block. (c) The ResNet feature extraction block, which includes a residual connection. (d) The SEResNet feature extraction block. Here, the Squeeze-and-Excitation block is further adopted in addition to the residual structure of (c). (e) The numbers of channels per stage of different grading models.

To validate the effectiveness of the proposed ENROL framework, we instantiate it with three typical architectures (VGG, ResNet, and SEResNet [27]-[29]) to handle 3D brain MR image data for glioma grading. The details of the developed models are given in Fig. 4. All three models have similar structures. The main difference lies in the design of the extraction block (shown in Fig. 4b, Fig. 4c, and Fig. 4d).

The developed grading models are trained using stochastic gradient descent (SGD) with momentum of 0.9. The initial learning rate is sampled from the set {1e-3, 1e-4, 5e-5, 1e-5}. We train the networks using a batch size of 16 for 300 epochs. As suggested by He et al. [30], we adopt Group Normalization (GN) instead of Batch Normalization (BN) to avoid the negative effect of the small batch size. The early-stopping

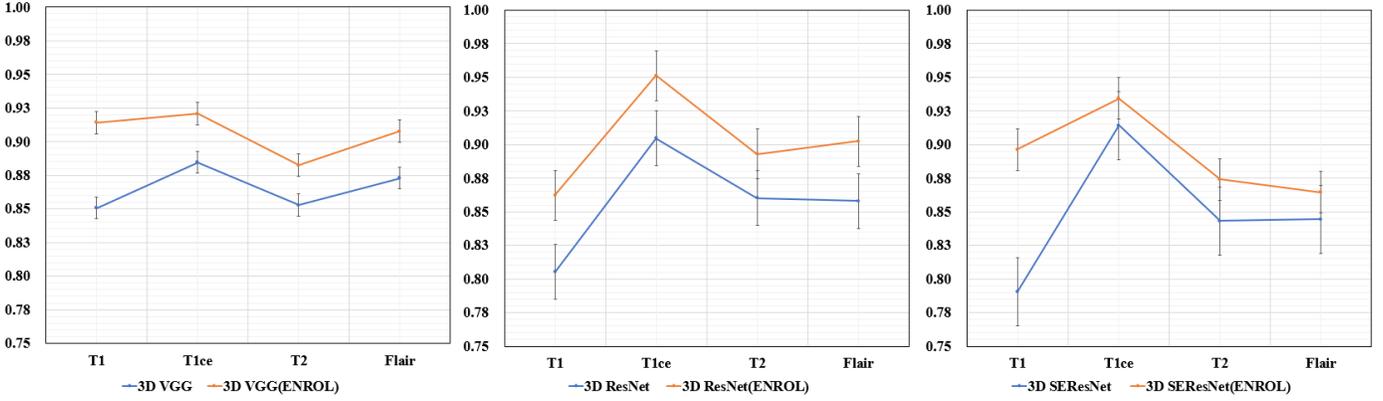

Fig. 5. AUC scores of the glioma grading models utilizing different network structures with or without our proposed ENROL framework. Compared to baseline models, the ENROL models can achieve better grading performances on three network structures. 3D ResNet trained with the ENROL framework gives the best results.

| Sequence | Basic Model | ENROL | AUC | Acc | Sen |
|---|---|---|---|---|---|
| T1 | 3D VGG | ✗ | 0.8506 | 0.7867 | 0.8983 |
| | | ✓ | **0.9142** | **0.8400** | **0.9661** |
| | 3D ResNet | ✗ | 0.8051 | 0.8267 | 0.9153 |
| | | ✓ | **0.8623** | **0.8400** | **0.9322** |
| | 3D SEResNet | ✗ | 0.7903 | 0.8133 | 0.9492 |
| | | ✓ | **0.8962** | **0.8267** | **0.9153** |
| T1ce | 3D VGG | ✗ | 0.8845 | 0.8400 | 0.9153 |
| | | ✓ | **0.9206** | **0.8533** | **0.9322** |
| | 3D ResNet | ✗ | 0.9047 | 0.8533 | 0.9322 |
| | | ✓ | **0.9513** | **0.8800** | **0.9831** |
| | 3D SEResNet | ✗ | 0.9142 | 0.8400 | 0.9153 |
| | | ✓ | **0.9343** | **0.8800** | **0.9492** |
| T2 | 3D VGG | ✗ | 0.8528 | 0.8133 | 0.9322 |
| | | ✓ | **0.8824** | **0.8400** | **0.9661** |
| | 3D ResNet | ✗ | 0.8602 | 0.8000 | 0.9492 |
| | | ✓ | **0.8930** | **0.8400** | **0.9661** |
| | 3D SEResNet | ✗ | 0.8432 | 0.7867 | 0.9831 |
| | | ✓ | **0.8739** | **0.8133** | **1.0000** |
| Flair | 3D VGG | ✗ | 0.8729 | 0.8000 | 0.9831 |
| | | ✓ | **0.9078** | **0.8267** | **1.0000** |
| | 3D ResNet | ✗ | 0.8581 | 0.7867 | 1.0000 |
| | | ✓ | **0.9025** | **0.7867** | **1.0000** |
| | 3D SEResNet | ✗ | 0.8443 | 0.7867 | 1.0000 |
| | | ✓ | **0.8644** | **0.8000** | **1.0000** |

Table. 1. Performance of the grading models trained with or without the proposed ENROL framework. The ENROL models can always obtain better performance

mechanism is implemented with a patience (the number of epochs to wait before early stop if no progress on the validation set is observed) of 20 to avoid serious overfitting. For fair comparisons, all three grading models are designed at the same scale and no additional regularization or data augmentation methods are included. Our experiments are run on two NVIDIA TITAN V GPUs, each with 12GB memory. All work related to deep learning is based on the open-source platform TensorFlow v2.5.1. The area under receiver operating characteristic curve (AUC) is utilized as the main evaluation metric. Accuracy (Acc) and sensitivity (Sen) are also calculated.

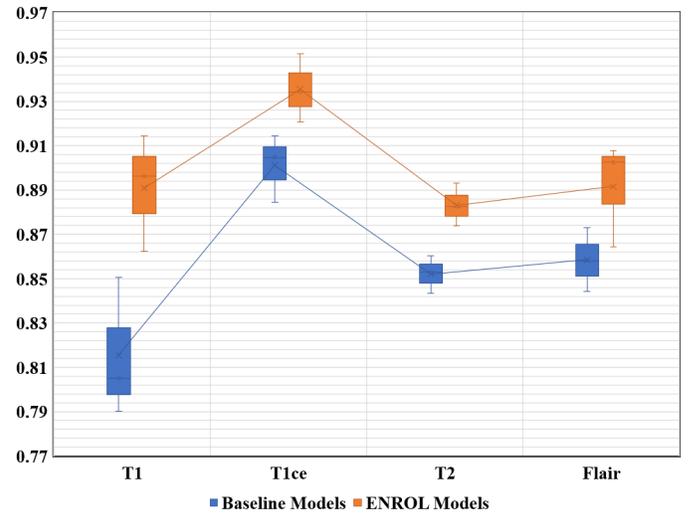

## III. RESULTS

### A. Performance Evaluation of the ENROL Framework

Glioma grading results of the models trained with or without the ENROL framework are listed in Table 1. Compared to the grading models trained without ENROL framework (baseline models), the ones trained with the ENROL framework (ENROL models) achieve obviously better grading performance. For all three network architectures (VGG, ResNet, and SEResNet), the proposed ENROL framework can always enhance the grading performance by achieving higher AUCs. The expert knowledge embedded in the quantitative hand-crafted features can indeed help the optimization of the grading models. The AUC scores of the grading models are plotted in Fig. 5 for direct comparison. The results suggest that 3D ResNet trained with the proposed ENROL framework on T1ce MR images gives the best grading performance.

Fig. 6. Improved glioma grading performances of the ENROL models on different MR sequences. Compared to the baseline models, the ENROL models can improve the grading performances significantly on all four MR sequences. Among the four sequences, T1ce MR images present the best grading potential.

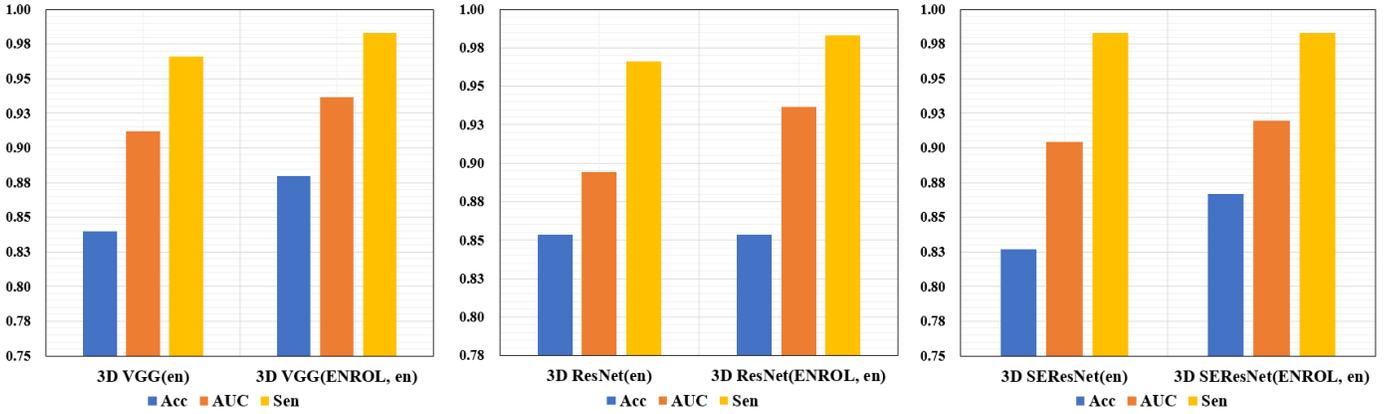

**Fig. 7.** Performances of the ensemble models. We ensemble the grading models trained with MR image data acquired with different sequences. Average ensembling is utilized. Results suggest that the ENROL framework can also improve the performance of the ensemble models.

### B. Performance Evaluation of Different MR Sequences

The performance of the developed grading models utilizing data acquired with different MR sequences are different. The AUC scores of the baseline models and the ENROL models on MR image data acquired with different sequences are plotted in Fig. 6. The average improvements of the AUC scores on four sequences (T1, T1ce, T2 and Flair) are $0.0756 \pm 0.0265$ ($p = 0.0331$), $0.0343 \pm 0.0133$ ($p = 0.0515$), $0.0310 \pm 0.0016$ ($p = 0.0138$) and $0.0331 \pm 0.0122$ ($p = 0.1067$) respectively. These results suggest that the proposed ENROL framework can improve the glioma grading performance significantly on all four MR sequences. Among the four sequences, T1ce MRI presents the best glioma grading potential. All the grading models achieve the highest performance on T1ce MR images. As a result, we suggested that glioma grading models built on T1ce MR images may serve as a more valuable baseline to verify the performance of the proposed grading model. Moreover, ensemble models are developed based on the models trained on four respective sequences. The ensemble models take the average predictions on four sequences as the final results. AUC scores of the ensemble models are plotted in Fig. 7. The results confirm that the introduced expert knowledge can also improve the performance of the ensemble models.

### C. Performance Evaluation of Different Architectures

Different network structures are investigated to further validate the effectiveness of the proposed ENROL framework. Among the three baseline models, 3D VGG achieves the best results. Similarly, with the proposed ENROL framework, 3D VGG is still better than 3D ResNet and 3D SEResNet for the glioma grading task. This phenomenon is quite counterintuitive. We speculate that it is related to the overfitting issue. For the three baseline models, 3D SEResNet contains the most nonlinear calculations with the residual block and sequence-and-extraction block and 3D VGG has the fewest with only the stack structures. More nonlinearity calculations empower the model with enhanced feature extraction and feature representation capabilities, but in the meantime, more training data are required for the model optimization. Although the ENROL models can use MR images data more effectively than the baseline models, the issue of insufficient training data for 3D ResNet still exists. Therefore, worse results are obtained for the more complex models. The grading results (AUCs and ACCs) of three architectures trained with the ENROL framework averaged over the four sequences are plotted in Fig. 8. These results suggest that excessive nonlinearity calculations of ResNet architecture and SEResNet architecture lead to worse convergence behaviors, whereas VGG architecture with relatively fewer nonlinearity calculations can get more stable and better grading performances on all four sequences.

### D. Performance Evaluation using Grading Models with or without the Segmentation Maps.

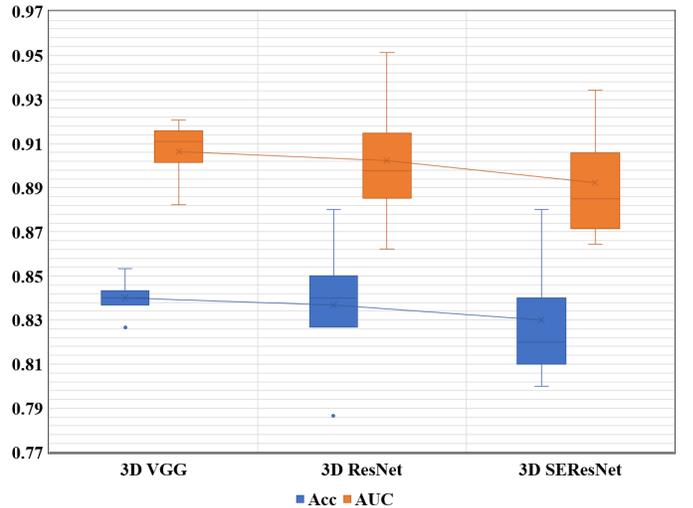

**Fig. 8.** Glioma grading performances of models utilizing different network structures. All three networks have similar parameter counts. ResNet and SEResNet have more nonlinearity calculations than VGG. Regardless of whether the ENROL framework is used for training, the excessive nonlinearity calculations lead to worse convergence behaviors, and VGG with relatively fewer nonlinearity calculations can achieve more stable and better glioma grading performance.

The ENROL models is physician-friendly, which can be tested without requiring the inputs of brain lesion segmentation maps. For comparisons, we also developed a 3D Seg-CNN and a radiomics-based model that require the lesion segmentation maps in both the training and the testing phases. The 3D Seg-CNN was developed based on VGG-like architecture for glioma grading. The lesion regions were cropped and resized to 64×64×64, and then, they were treated as the inputs to train the 3D Seg-CNN. For the radiomics-based model, we used the hand-crafted features extracted in the expert knowledge construction stage to directly develop a radiomics classifier for glioma grading. The key features were selected utilizing a three-level cascaded feature selection algorithm (student t-test, lasso, and recursive feature elimination algorithm). We trained a logistic regression classifier with the selected features, which was the final glioma grading model.

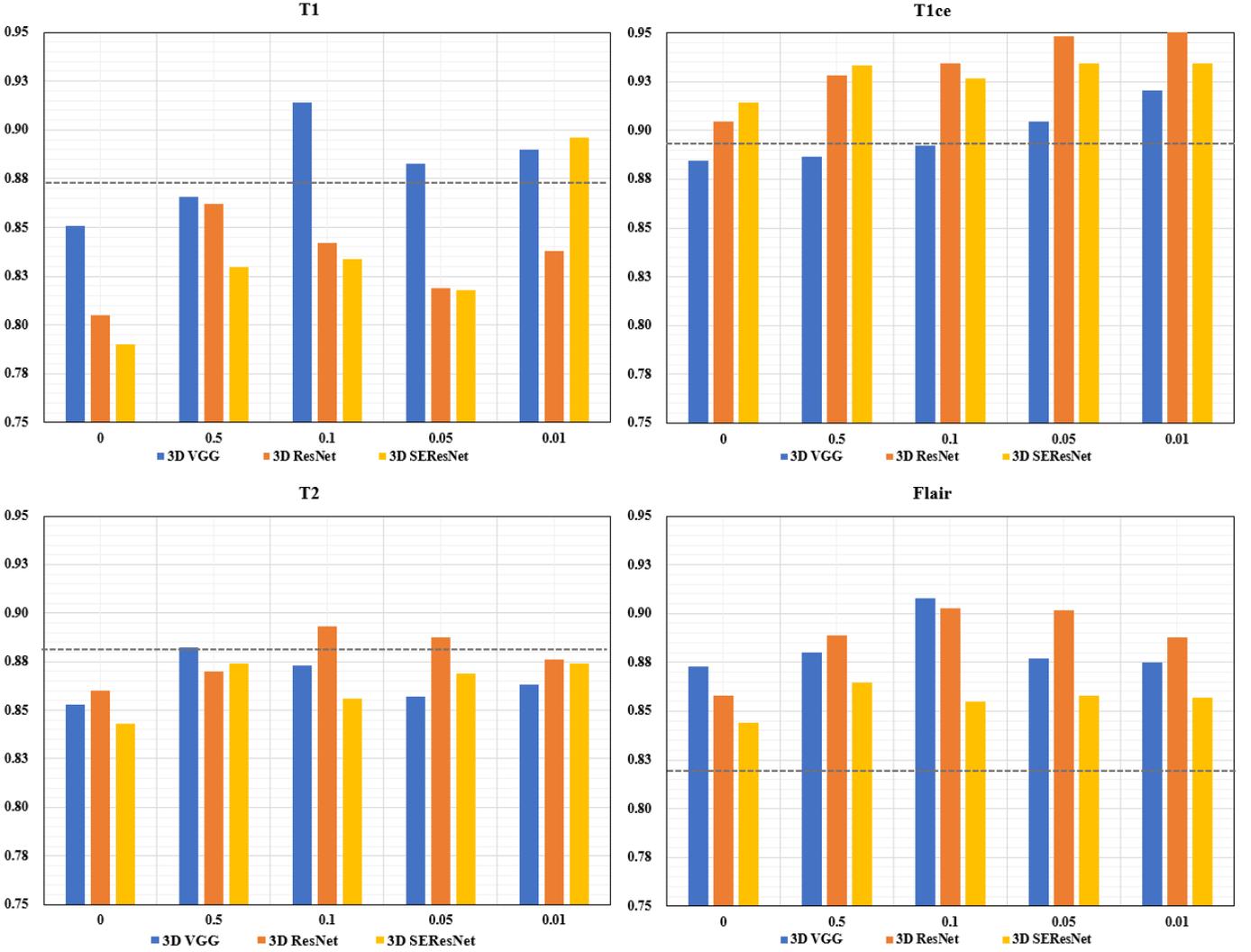

**Fig. 9.** AUC scores of the ENROL models with different values of $\lambda$. We test the influence of the hyperparameters $\lambda$ in the ENROL framework on the performance of the grading models. We also show the grading performance of radiomic models with the gray line in each plot.

| Model | Need Segmentation maps in training | Need Segmentation Maps in testing | AUC | Acc | Sen |
|---|---|---|---|---|---|
| 3D VGG | ✗ | ✗ | 0.8845 | 0.8400 | 0.9153 |
| 3D ResNet | ✗ | ✗ | 0.9047 | **0.8533** | **0.9322** |
| 3D SEResNet | ✗ | ✗ | **0.9142** | 0.8400 | 0.9153 |
| 3D Seg-CNN | ✓ | ✓ | **0.9025** | 0.8533 | **1.0000** |
| Radiomic Model | ✓ | ✓ | 0.8930 | **0.8933** | 0.9492 |
| 3D VGG(ENROL) | ✓ | ✗ | 0.9206 | 0.8533 | 0.9322 |
| 3D ResNet(ENROL) | ✓ | ✗ | **0.9513** | **0.8800** | **0.9831** |
| 3D SEResNet(ENROL) | ✓ | ✗ | 0.9343 | 0.8800 | 0.9492 |

**Table. 2.** Glioma grading performance of models trained and tested with or without the brain lesion segmentation maps. Our proposed ENROL framework can help models obtain promising grading performance without the cumbersome brain lesion segmentation procedure in the testing phase.

Results are listed in Table 2. It can be observed that using the lesion regions as inputs can indeed improve the grading performance. As a classical method for medical image analysis, the radiomics-based model can also generate promising results. On the other hand, compared to the 3D Seg-CNN and the radiomics-based model, our proposed ENROL framework can help models obtain similar or even better grading performance without the cumbersome brain lesion segmentation procedure in the testing phase.

### E. Performance Evaluation of the Hyperparameters in the ENROL Framework

We studied the influence of the parameters λ in the ENROL framework on the performance of the grading models. The three different network architectures (VGG, ResNet and SEResNet) were tested on four sequences (T1, T1ce, T2 and Flair). The parameter λ is sampled from the set {0, 0.5, 0.1, 0.05, 0.01}. Here, the ENROL framework with λ = 0 means that the grading models is trained as baseline models. In addition, we also trained the radiomic models on four sequences, and the training process of radiomic models is mentioned above. The AUC scores are plotted in Fig 9. As the results suggested, the proposed ENROL framework can always improve the performance of the grading models on four sequences, and the ENROL models can be close to or better that radiomic models in each sequence.

## IV. CONCLUSION

In this study, we propose an expert knowledge-guided geometric representation learning framework to achieve segmentation-free CNN-based glioma grading utilizing MR images. Geometric representation learning is introduced to effectively fuse the non-Euclidean hand-crafted features with the non-Euclidean learned features extracted from grading MR images. A manifold discrepancy distance is designed to exploit the expert knowledge embedded in the hand-crafted features for the parameter optimization of the grading model. In this way, promising glioma grading performance is obtained without requiring brain lesion segmentation maps at the testing phase. The proposed ENROL framework is flexible and physician-friendly, which has a high potential to be utilized as a tool to assist physicians in achieving fast and accurate glioma grading.